\documentclass[12pt,dvips]{article}
\usepackage[dvips]{graphicx}
\usepackage{graphicx}
\usepackage{latexsym}
\usepackage[nosort]{cite}
\parskip=\itemsep               
\graphicspath{{plots/}}
\DeclareGraphicsExtensions{.eps.gz,.eps,.ps,.ps.gz}
\newcommand{\lwig}{\mbox{\,\raisebox{.3ex}
    {$<$}$\!\!\!\!\!$\raisebox{-.9ex}{$\sim$}\,}}
\newcommand{\gwig}{\mbox{\,\raisebox{.3ex}
    {$>$}$\!\!\!\!\!$\raisebox{-.9ex}{$\sim$}}\,}
\newcommand{\rav}{\langle\rho\rangle}
\newcommand{\xbj}{x_{\rm Bj}}

\newcommand{\xpr}{{x^\prime}}
\newcommand{\Qpr}{Q^{\prime 2}}

\newcommand{\rb}[1]{\raisebox{2.5ex}[-1.5ex]{#1}}

\def\funp{{I\!\!P}}
\def\Journal#1#2#3#4{{#1} {\bf #2} (#4) #3}

\def\NPA{{ Nucl. Phys.} A}
\def\NPB{{ Nucl. Phys.} B}
\def\PLB{{ Phys. Lett.}  B}
\def\PRL{ Phys. Rev. Lett.}
\def\PRD{{ Phys. Rev.} D}
\def\PR{Phys. Rep.}
\def\ZPC{{ Z. Phys.} C}

\def\CPC{ Comput. Phys. Commun.}

\def\JPG{ J. Phys. G }
\def\MPL{ Mod. Phys. Lett. }

\date{\empty}

\title{{\normalsize\rightline{DESY 02-094}\rightline{hep-ph/0207300}}
\vskip 1cm 
 \bf QCD Instantons and High-Energy Diffractive Scattering
       \vspace{21mm}} 
\author{F. Schrempp and A. Utermann\\[4mm] 
Deutsches Elektronen-Synchrotron DESY, Hamburg, Germany}
\begin{document}
\begin{titlepage} 
  \maketitle
\vspace{3cm}
\begin{abstract}
We pursue the intriguing possibility that
larger-size instantons build up  diffractive
scattering, with the marked instanton-size scale $\rav\approx 0.5$ 
fm being reflected in the conspicuous ``geometrization'' of soft QCD.
As an explicit step in this direction, the known instanton-induced
cross sections in deep-inelastic scattering (DIS) are transformed into the
familiar colour dipole picture, which represents an intuitive
framework for investigating the transition from hard to soft physics
in DIS at small $x_{\rm Bj}$. The simplest instanton ($I$) process without
final-state gluons is studied first. With the help of lattice results,  
the $q\overline{q}$-dipole size $r$ is carefully increased towards hadronic
dimensions. Unlike perturbative QCD, one now observes a competition between two
crucial length scales: the dipole size $r$ and the size $\rho$ of the
background instanton that is sharply localized around $\rav\approx
0.5$ fm. For $r\,\gwig\,\rav$, the dipole cross section indeed
saturates towards a geometrical limit, proportional to the area
$\pi\,\rav^2$, subtended by the instanton. In case of final-state
gluons, lattice data are crucially used to support the emerging
picture and to assert the range of validity of the underlying $I\bar{I}$-valley
approach. As function of an appropriate energy variable, the
resulting dipole cross section turns out to be sharply peaked at the
sphaleron mass in the soft regime. The general geometrical features
remain like in the case without gluons.
\end{abstract}

\thispagestyle{empty}
\end{titlepage}
\newpage \setcounter{page}{2}

\begin{figure} 
\begin{center}
\parbox{5.2cm}{\includegraphics*[width=5.2cm]{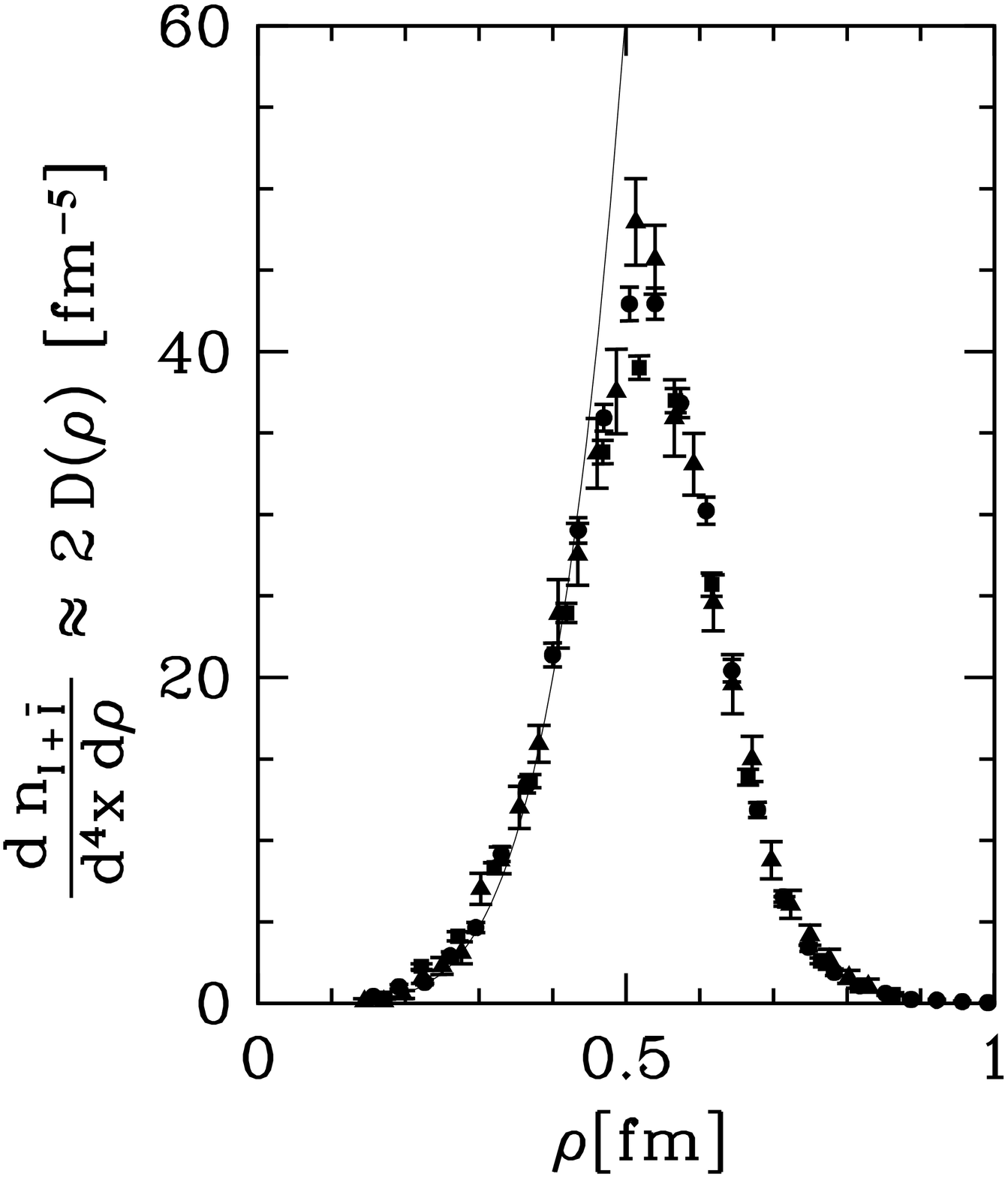}}\hfill
\parbox{11cm}{\includegraphics*[width=11cm]{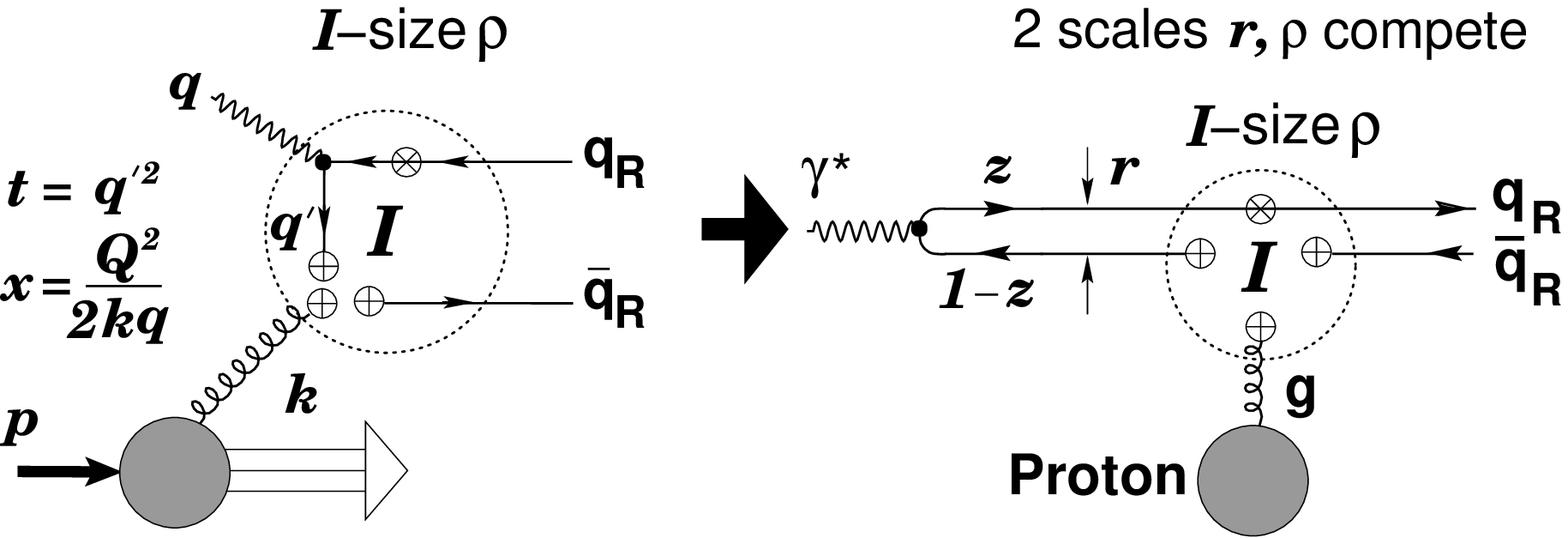}}
 \caption[dum]{\small\label{pic} (Left) UKQCD
 lattice data~\cite{ukqcd,rs-lat,rs3} of the 
 $(I+\bar{I})$-size distribution for quenched QCD ($n_f = 0)$. Both
 the sharply defined $I$-size scale 
$\langle\rho\rangle \approx 0.5$ fm and the parameter-free agreement with
\mbox{$I$-perturbation} theory~\cite{rs-lat,rs3} for $\rho\lwig 0.35$ fm are
apparent (solid line $\Leftrightarrow$ Eq.~(\ref{density}) with 3-loop
expression of $\alpha_{\rm s}$ and $\Lambda^{(\rm n_f =0)}_{\rm
\overline{MS}}=238$ MeV~\cite{alpha}).
(Right) Transcription of the simplest
 \mbox{$I$-induced} process ($n_f = 1,\ n_g = 0$)  with variables $x$ and $t$
 into the colour dipole picture with the variables $z$ and $\mathbf r$}
\end{center}
\end{figure}

{\em 1.} QCD instantons~\cite{bpst} are non-perturbative
fluctuations of the gluon fields, with a size distribution {\em sharply}
localized around $\rav\approx 0.5~{\rm fm}$ according to lattice
simulations~\cite{ukqcd} (Fig. \ref{pic} (left)). They are well known
to induce chirality-violating processes, absent in conventional perturbation
theory~\cite{th}.  Deep-inelastic
scattering\footnote{For an exploratory calculation of
the instanton contribution to the gluon structure function, see
Ref.~\cite{bb}} (DIS) at HERA has been shown to offer a unique
opportunity~\cite{rs1} for discovering such processes
induced by {\it small} instantons ($I$) through a sizeable
rate~\cite{mrs,rs2,rs-lat} and a characteristic final-state
signature~\cite{rs1,qcdins,rs3}. An intriguing but non-conclusive excess
of events in an ``instanton-sensitive'' data sample, has recently been
reported in the first dedicated search for instanton-induced processes
in DIS at HERA~\cite{h1-final}.

The validity of $I$-perturbation
theory in DIS is warranted by some (generic) hard momentum scale
$\mathcal{Q}$ that ensures a dynamical
suppression~\cite{mrs} of contributions from larger size instantons
with $\rho\gwig \mathcal{O}(1/\mathcal{Q})$. Here, the above mentioned
intrinsic instanton-size scale $\rav\approx 0.5~{\rm fm}$ is correspondingly
unimportant.    

This paper, in contrast, is devoted to the intriguing question about
the r{\^o}le of {\em larger-size} instantons and the associated intrinsic scale
\mbox{$\rav\approx 0.5~{\rm fm}$}, for decreasing ($Q^2,\,\xbj$)
towards the~soft scattering regime. A number of authors have focused
attention recently  
on the interesting possibility that larger-size instantons may well be
associated with a dominant part of soft high-energy scattering, or even make up
diffractive scattering
altogether~\cite{levin,shuryak1,shuryak11,fs,shuryak2,su1}. We shall
argue below that the instanton scale $\rav$ is reflected in the
conspicuous {\it geometrization} of soft QCD.  

There are two immediate qualitative reasons for this idea. 
First of all, instantons represent truly non-perturbative gluons that
naturally bring in an intrinsic size scale $\rav\approx 0.5~{\rm fm}$ 
of hadronic dimension (Fig.~\ref{pic} (left)). The instanton size 
happens to be surprisingly close to a corresponding ``diffractive'' size
scale, $R_\funp =\,R\,\sqrt{\alpha^\prime_\funp/\alpha^\prime}
\approx 0.5$ fm, resulting from simple dimensional rescaling along
with a generic hadronic size \mbox{$R\approx 1$ fm} and the abnormally small $\funp$omeron slope
$\alpha^\prime_\funp\approx \frac{1}{4}\,\alpha^\prime$ in terms of the normal,
universal Regge slope  $\alpha^\prime$. 

Secondly, we know already from $I$-perturbation theory that the instanton
contribution tends to strongly increase towards the infrared
regime~\cite{rs1,rs2,qcdins}. The mechanism for the decreasing
instanton suppression with increasing energy is known since a long
time~\cite{sphal2,shuryak2}: Feeding increasing energy into the scattering
process makes the picture shift from one 
of tunneling between vacua ($E\approx 0$) to that of the actual
creation of the sphaleron-like  configuration~\cite{sphal1} on top of
the potential barrier of height~\cite{rs1} $E = M_{\rm
sph}\propto\frac{1}{\alpha_s\rho_{\rm eff.}}$. In a second step,
the action is real and the sphaleron then decays into a multi-parton
final state.  

The familiar colour dipole picture~\cite{dipole} represents a
convenient and intuitive framework for investigating the transition from
hard to soft physics (diffraction) in DIS at small $\xbj$. 
At the same time, this picture is very well suited for studying the crucial
interplay between the $q\overline{q}$-dipole size $r$ 
and the instanton size $\rho$ in an explicit and well-defined manner, as we
shall discuss next. 

The intuitive content of the colour dipole picture is that at high
energies, in the  proton's rest frame, the virtual photon fluctuates
predominantly into a $q\overline{q}$-dipole a long distance
upstream of the target proton.
The large difference of the $\gamma^\ast\rightarrow
q\overline{q}$-dipole formation and 
$(q\overline{q})$-$P$ interaction times in the proton's rest frame at
small $\xbj$ then generically gives rise to the familiar factorized expression of
the inclusive photon-proton cross sections, 
\begin{equation}
\sigma_{L,T}(\xbj,Q^2) 
=\int_0^1 d z \int d^2\mathbf{r}\; |\Psi_{L,T}(z,r)|^2\,\sigma_{\rm
dipole}(r,\ldots),  
\label{dipole-cross}
\end{equation}
in terms of the modulus squared of the 
(light-cone) wave function\footnote{While quark mass effects are known
to become important at the larger distances of interest here, these are
hard to explicitly account for in the instanton-calculus and thus
beyond the scope of the present paper.} 
of the virtual photon, calculable in pQCD
($\hat{Q}=\sqrt{z\,(1-z)}\,Q;\ r=\mid{\mathbf r}\mid$), 
\begin{equation}
\label{wavefu}
 \mid\Psi^{\rm pQCD}_{ L\atop T}(z,r)\mid^{\,2}
 =\sum_q e_q^2\,\frac{6\alpha_{\rm em}\,\hat{Q}^2}{(2\pi)^2}\,{\rm
 K}_{0\atop
 1}(\hat{Q}\,r)^2\left\{\begin{array}{l}4\,z\,(1-z)\\  
(z^2+(1-z)^2)\,,\end{array}\right.
\end{equation}  
and the $q\overline{q}$-dipole\,-\,nucleon cross section $\sigma_{\rm
dipole}(r,\ldots)$. The variables in Eq.~(\ref{dipole-cross}) denote
the transverse $(q\overline{q})$-size $\mathbf r $ 
and the photon's longitudinal momentum fraction $z$ carried by the quark. 
$\Psi_{L,T}(z,r)$ contains the
dependence on the $\gamma^\ast$-helicity. The dipole cross section is
expected to include in general the main non-perturbative contributions. 
For small $r$, however, one finds within
pQCD~\cite{dipole,dipole-pqcd} that $\sigma_{\rm dipole}$ vanishes 
with the area $\pi r^2$ of the $q\overline{q}$-dipole. Besides this phenomenon
of ``colour transparency'' for small $r$,  the dipole cross
section is expected to saturate towards a constant, once the
$q\overline{q}$-separation $r$ has reached hadronic distances. 
$$
\sigma_{\rm dipole}\ \left\{\begin{array}{lrcl}
\sim & \pi\,r^{\,2},&
{r}^2\lwig\mathcal{O}(\frac{1}{Q^2}),&\mbox{\rm\ ``colour
transparency''~\cite{dipole,dipole-pqcd}},\\[1ex] 
\approx &{\rm constant},&r\,\gwig\, 0.5 {\rm\ fm},& \mbox{\rm\
``hadron-like,\ saturation''.}\end{array}\right.
$$ 
The strategy is now to transform the known results on
$I$-induced processes in DIS  into this intuitive colour dipole
picture. We shall begin with the most transparent case of the simplest
$I$-induced process~\cite{mrs},
\begin{equation}
\gamma^\ast\,g\stackrel{(I)}{\Rightarrow} q_{\rm R}\
\overline{q}_{\rm R}, 
\label{simplest}
\end{equation}
for one flavour  and no final-state
gluons. Subsequently, we shall turn to the more realistic case~\cite{rs2} with
final-state gluons and $n_f\ (=3)$ light flavours.

The idea is to consider first large $Q^2$ and appropriate cuts on the
variables $z$ and $r$, such
that \mbox{$I$-per}turbation theory holds. By exploiting the lattice results
on the instanton-size distribution (Fig.~\ref{pic} (left)),
we shall then carefully increase the $q\overline{q}$-dipole size $r$ towards
hadronic dimensions. 


{\em 2.} Let us start by recalling the relevant results~\cite{mrs} for the
simplest $I$-induced DIS process
(\ref{simplest}), corresponding to one flavour ($n_f=1$) and no
final-state gluons (Fig.~\ref{pic} right). At small $\xbj
=\frac{Q^2}{2\,P\cdot q}$, the leading $I$-induced contribution to the
respective partonic cross sections comes from the $\gamma^\ast g$
subprocess. In terms of the gluon density $G(\xbj,\mu^2)$, the results
from Ref.~\cite{mrs} for the $\gamma^\ast N$ cross sections $\sigma_T
(\xbj ,Q^2)$ and $\sigma_L (\xbj ,Q^2 )$ for transverse ($T$) and
longitudinal ($L$) virtual photons, respectively, then take the
following form,  
\begin{eqnarray}
\sigma_{L,T}(\xbj,Q^2)&=&
\int_{\xbj}^1 \frac{d x}{x}\left(\frac{\xbj}{
x}\right)G\left(\frac{\xbj}{x},\mu^2\right)\int d  t\, \frac{d
\hat{\sigma}_{L,T}^{\gamma^* g}(x,t,Q^2)}{d t};\,\label{general}\\[2ex] 
\frac{d\hat{\sigma}_{L}^{\gamma^* g}}{d  t}&=&\frac{\pi^7}{2}
\frac{e_q^2}{Q^2}\frac{\alpha_{\rm em}}{\alpha_{\rm
s}}\left[x(1-x)\sqrt{t u}\,  \frac{\mathcal{R}(\sqrt{-
t})-\mathcal{R}(Q)}{t+Q^2}-(t\leftrightarrow  u)\right]^{\,2} \label{mrsL}\\ 
\frac{d\hat{\sigma}_{T}^{\gamma^* g}}{d  t}&=&\frac{\pi^7}{8}
\frac{e_q^2}{Q^2}\frac{\alpha_{\rm em}}{\alpha_{\rm s}}x(1-x)\left\{\left[\mathcal{R}(\sqrt{-t})^{\,2}+tu\left(\frac{\mathcal{R}(\sqrt{-t})-\mathcal{R}(Q)}{t+Q^2}\right)^{\, 2}+(t\leftrightarrow u)\right]\right.    \nonumber \\ &&
\left.+tu\left[\frac{\mathcal{R}(\sqrt{-t})-\mathcal{R}(Q)}{t+Q^2}-(t\leftrightarrow  u)\right]^{\,2}(2x(1-x)-1)\right\}.\label{mrsT} 
\end{eqnarray}
Eqs.~(\ref{mrsL}),~(\ref{mrsT}) involve the master integral
$\mathcal{R}(\mathcal{Q})$ with dimensions of a length,  
\begin{equation}
\mathcal{R}(\mathcal{Q})=\int_0^{\infty} d\rho\,D(\rho)\rho^5(\mathcal{Q}\rho)\mbox{K}_1(\mathcal{Q}\rho).
\label{masterI}
\end{equation}
The $I$-size distribution $D(\rho)$ enters in Eq.~(\ref{masterI}) as a
crucial building block of the 
$I$-calculus. For small $\rho$ (probed at large $\mathcal{Q}$)
$D(\rho)$ is explicitly known within $I$-perturbation theory
\cite{th,morretal}. Correspondingly, in Ref.~\cite{mrs}, the integral
(\ref{masterI}) was carried out explicitly by specializing on the
familiar $I$-perturbative form (renormalization scale $\mu_r$),  
\begin{eqnarray}
   D(\rho)\,\Rightarrow\, D^{(I)}_{I-{\rm pert}}(\rho)=D^{(\bar{I})}_{I-{\rm pert}}(\rho)
   &=&
   \frac{d_{\rm \overline{MS}}}{\rho^{\,5}}
   \left(\frac{2\pi}{\alpha_{\rm \overline{MS}}(\mu_r)}\right)^6 
   \exp{\left(-\frac{2\pi}{\alpha_{\rm \overline{MS}}(\mu_r)}\right)}(\rho\,
   \mu_r)^b,  
   \label{density}
\\ 
\label{b-morretal}
b&=&\beta_0+(\beta_1-12\,\beta_0)\frac{\alpha_{\rm
   \overline{MS}}(\mu_r)}{4\,\pi}
\end{eqnarray}  
in terms of the QCD $\beta$-function coefficients,
$\beta_0=11-\frac{2}{3}{n_f},\  \beta_1=102-\frac{38}{3} {n_f}$ and
the known, scheme-dependent constant $d_{\rm \overline{MS}}=C_1\exp[-3
C_2+n_f C_3]/2$ with $C1=0.46628,\ C_2=1.51137$, and $C_3=0.29175$. In
this form, it satisfies renormalization-group invariance at the
two-loop level~\cite{morretal},
i.e. $D^{-1}_{\,I-{\rm pert}}\,d\,D_{\,I-{\rm pert}}/d\,\ln(\mu_r)=\mathcal{O}(\alpha_{\rm s})^2$.  

In this paper we prefer to adopt a more general attitude concerning
the form of $D(\rho)$ and thus leave the integral
(\ref{masterI}) unevaluated for the time being. 
For larger $I$-size $\rho$ (as relevant for \mbox{smaller $\mathcal{Q}$}), $D(\rho)$ is known from lattice simulations
(Fig.~\ref{pic} (left)). A striking feature is the strong peaking of
$D_{\rm lattice}(\rho)$ around $\langle\rho\rangle\approx 0.5$~fm, whence 
$\mathcal{R}(0)=\int_0^{\infty} d\rho\;D_{\rm lattice}(\rho)\rho^5$
is finite. For $D_{\rm lattice}(\rho)$ from Fig.~\ref{pic} (left),
one finds $\mathcal{R}(0)$ to be numerically close\footnote{More quantitatively,
it is usually the peak position $\rho=\rho_{\rm peak}\approx 0.59$ fm of
$\rho^5\,D_{\rm lattice}(\rho)$ that sets the scale. For simplicity,
we shall mostly ignore here the slight numerical difference between
$\rho_{\rm peak}$ and $\langle\rho\rangle\approx 0.51$ fm} to $\langle\rho\rangle$.

By means of an appropriate change of variables 
and a subsequent $2d$-Fourier transformation,
Eqs.~(\ref{general}) - (\ref{mrsT}) may 
indeed be cast into a colour dipole form, 
\vspace{-1ex}
\begin{equation}
\sigma_{L,T}=
\int_{\xbj}^1 \frac{d x}{x}
\int d t\, \frac{d \hat{\sigma}_{L,T}^{\gamma^* g}}{d
t}\,\{\ldots\}\Rightarrow \int dz\int d^2\mathbf{r} \,
\left(|\Psi_{L,T}|^2 \sigma_{\rm dipole}\right)^{(I)}. 
\label{siglt}
\end{equation}
The change of variables used is
$(t,\ x)\Rightarrow (\mathbf{l}^2, z)$, with $\mathbf{l}$ being the
quark transverse
momentum and $z$ the photon's longitudinal momentum
fraction carried by the quark, 
\begin{equation}
\left.\mbox{$\begin{array}{rclclcl}
-t&=&Q^{\prime\,2}&=&\frac{\hat{Q}^2+l^2}{z};\hspace{2ex}
-u&=&\frac{\hat{Q}^2+l^2}{1-z}\\[1ex]
x&&\phantom{Q^{\prime\,2}}&=&\frac{\hat{Q}^2}{\hat{Q}^2+l^2};
\end{array}$} \right\}\hspace{2ex}
\hat{Q}=\sqrt{z\,(1-z)}\,Q;\hspace{3ex} l=\mid\mathbf{l}\mid.
\label{vartrans}
\end{equation}
The subsequent $2d$-Fourier transformation then introduces 
the transverse $q\overline{q}$ distance
$\mathbf r$ of the colour-dipole picture via
\begin{eqnarray}
G(r,\ldots)
&=&\int\frac{d^2\mathbf{l}}{(2\,\pi)^2}\,e^{i\mathbf{r}\cdot
\mathbf{l}}\,\tilde{G}(l,\ldots)=\frac{1}{2\,\pi}\,\int dl\,l\, {\rm
J}_0(l\,r)\,\tilde{G}(l,\ldots);\hspace{2ex}\mbox{\rm \ and}\label{2dFT1}\\
\int \frac{d^2\mathbf{l}}{(2\pi)^2}\,\tilde{G}(l,\ldots)^2 &=& \int d^2\mathbf{r}\,G(r,\ldots)^2;\hspace{1ex} 
\int \frac{d^2\mathbf{l}}{(2\,\pi)^2}\,l^2
\,\tilde{G}(l,\ldots)^2 = \int d^2\mathbf{r}\,\left(\frac{d}{dr}G(r,\ldots)\right)^2.
\label{2dFT2}
\end{eqnarray}
Like is usual in pQCD-calculations~\cite{dipole-pqcd}, we throughout
invoke the familiar ``leading-$\log(1/\xbj)$'' - approximation, 
$\xbj/x\,G(\xbj/x,\mu^2) \approx \xbj G(\xbj,\mu^2)$, 
for simplicity.   
In terms of the familiar pQCD wave function (\ref{wavefu}) of the
photon, we then obtain from Eqs.~(\ref{general}) - (\ref{mrsT}) the
following integrands on the r.h.s. of Eqs.~(\ref{siglt}), 
\begin{eqnarray}
\lefteqn{
\left(\left|\Psi_L\right|^2\sigma_{\rm dipole}\right)^{(I)}
 \approx\,\mid\Psi_L^{\rm pQCD}(z,r)\mid^{\,2}\,
\frac{1}{\alpha_{\rm s}}\,\xbj\, G(\xbj,\mu^2)\,\frac{\pi^8}{12}}\nonumber\\[1ex]
&&\times\left\{\int_0^\infty\,d\rho\, D(\rho)\,\rho^5\,\left(\frac{-\frac{d}{dr^2}\left(2 r^2
\frac{\mbox{K}_1(\hat{Q}\sqrt{r^2+\rho^2/z})}{\hat{Q}\sqrt{r^2+\rho^2/z}}\right)}{{\rm K}_0(\hat{Q}r)}-(z\leftrightarrow 1-z) \right) 
\right\}^2, \label{resultL}\\
\lefteqn{
\left(\left|\Psi_T\right|^2\sigma_{\rm dipole}\right)^{(I)}
 \approx\,\mid\Psi_T^{\rm pQCD}(z,r)\mid^{\,2}\,
\frac{1}{\alpha_{\rm s}}\,\xbj\, G(\xbj,\mu^2)\,\frac{\pi^8}{12}}\nonumber\\[1ex]
&&\times\left\{\left(\int_0^\infty\,d\rho\, D(\rho)\,\rho^5\,\frac{
\frac{r\,\mbox{K}_1(\hat{Q}\sqrt{r^2+\rho^2/z})}{\sqrt{r^2+\rho^2/z}}}{{\rm K}_1(\hat{Q}r)\sqrt{z^2+(1-z)^2}}
\right)^{\,2}+(z\leftrightarrow 1-z)+\ldots \right\}.\label{resultT} 
\end{eqnarray}
As expected, one explicitly observes a 
{\em competition} between two crucial length scales in 
Eqs.~(\ref{resultL}), (\ref{resultT}): the size $r$ of the 
$q\overline{q}$-dipole and the typical size of the background
instanton of about $\rav\approx 0.5~{\rm fm}$. 
Like in pQCD, the asymmetric configuration, $z \gg
1-z$ or $1-z \gg z$, obviously dominates.

The validity of strict $I$-perturbation theory ($D(\rho)\equiv D_{I-{\rm
pert}}(\rho)$ in Eq.~(\ref{masterI})) requires the presence of a hard
scale $\mathcal{Q}$ along with certain cuts. 
However, after replacing $D(\rho)$ by $D_{\rm lattice}(\rho)$
(Fig.~\ref{pic} (left)), 
these restrictions are at least no longer  
necessary for reasons of convergence of the $\rho$-integral
(\ref{masterI}) etc.,  and one may tentatively 
increase the dipole size $r$ towards hadronic dimensions. 

 Next, we note in Eqs.~(\ref{resultL}), (\ref{resultT}), 
\vspace{-1ex}
\begin{eqnarray}
-\frac{d}{d\,r^2}\left(2\,r^2\frac{{\rm K}_1\left(\hat{Q}\sqrt{r^2+\rho^2/z}\right)}{\hat{Q}\sqrt{r^2+\rho^2/z}}\right) &\approx& \left\{\begin{array}{rl}
-\frac{{\rm K}_1\left(Q\, \rho\sqrt{1-z}\right)}{Q\,\rho\sqrt{1-z}}&\frac{r^2\,z}{\rho^2}\Rightarrow 0,\\[2ex]
{\rm K}_0\left(\hat{Q}\,r\right)&\frac{r^2\,z}{\rho^2}\mbox{\ large}.  
\end{array}\right.\label{approxL}\\
r\frac{{\rm
K}_1\left(\hat{Q}\sqrt{r^2+\rho^2/z}\right)}{\sqrt{r^2+\rho^2/z}}
&\approx& \left\{\begin{array}{rl}
\phantom{-\frac{{\rm K}_1\left(Q\, \rho\sqrt{1-z}\right)}{Q\,\rho\sqrt{1-z}}}&\phantom{\frac{r^2\,z}{\rho^2}\Rightarrow 0,}\vspace{-3ex}\\
\mathcal{O}(\frac{r\sqrt{z}}{\rho})&\frac{r^2\,z}{\rho^2}\Rightarrow 0,\\[2ex]
{\rm K}_1\left(\hat{Q}\,r\right)&\frac{r^2\,z}{\rho^2}\mbox{\ large}.  
\end{array}\right.
\label{approxT}
\end{eqnarray}
Due to the strong peaking of $D_{\rm
lattice}(\rho)$ around \mbox{$\rho\approx\rav$}, one finds from 
Eqs.~(\ref{resultL}) - (\ref{approxT}) for
the limiting cases of interest ($z\gg 1-z$ without restriction),
\begin{equation}
\begin{array}{|c|c|}
\hline 
r&\rule[-2mm]{0mm}{9mm}\left(\mid\Psi_{L,T}\mid^{\,2} \sigma_{\rm dipole}\right)^{(I)}\\[1ex]\hline
\rule[2mm]{0mm}{4mm}
\Rightarrow 0&\mathcal{O}(1),\mbox{\rm \ but\ exponentially\ small\
for\ large}\ \hat{Q},\\[3ex]
&\rule[-2mm]{0mm}{7mm} \mid\Psi^{\rm
 \,pQCD}_{L,T}\mid^{\,2}\, \sigma_{\rm 
 dipole}^{(I)}\hspace{2ex}\mbox{\rm\ with}\\
\rb{$\gwig \rav$}
&\displaystyle\rule[-3mm]{0mm}{11mm}\sigma^{(I)}_{\rm
dipole}(r,\ldots)=\frac{1}{\alpha_{\rm s}}\,\xbj\,G(\xbj,\mu^2)\,\frac{\pi^8}{12}\,\left(\int_0^\infty\,d\rho\,D_{\rm lattice}(\rho)\,\rho^5\right)^2.  
\\[3ex] \hline
\end{array}
\label{final}
\end{equation}
{\em In summary:} As apparent in Eqs.~(\ref{resultL}), (\ref{resultT}),
(\ref{final}), the dipole cross section from the simplest
\mbox{$I$-induced} process   
raises strongly around the instanton scale, $r\approx\langle\rho\rangle$, and 
indeed {\em saturates} for large $r/\rav$  towards a {\em constant
geometrical limit}, proportional to the area
$\pi\,\mathcal{R}(0)^2\, =\,  \pi\left(\int_0^\infty\,d\rho\,D_{\rm
lattice}(\rho)\,\rho^5\right)^2$, subtended by the instanton. Clearly,
without the crucial information about $D(\rho)$ from the 
lattice (Fig.~\ref{pic} (left)), the result would be infinite. Note the
inverse power of $\alpha_{\rm s}$ in front of $\sigma_{\rm  dipole}^{(I)}$
in Eq.~(\ref{final}), signalling its non-perturbative
nature\footnote{While the appropriate argument of
$\alpha_{\rm s}(\mu)$ is not quite obvious, a good guess might be $\mu\sim 1/\langle\rho\rangle$.}.

{\em 3.} We are now ready to turn to the more realistic $I$-induced
inclusive process
\begin{equation}
\gamma^\ast + g \stackrel{(I)}{\Rightarrow}
n_f\,(q_{\rm R} + \overline{q}_{\rm R}) + \mbox{\rm \ gluons}.
\label{dis}
\end{equation}
The corresponding DIS cross sections have been previously worked out
in detail~\cite{rs2} and are 
implemented in the Monte-Carlo generator QCDINS~\cite{qcdins} that
forms a basic tool in experimental searches for $I$-induced events at
HERA~\cite{h1-final}. 

The differential cross sections\footnote{Ignoring as usual non-planar
contributions~\cite{mrs,qcdins,rs3} that presumably are small
throughout most of the 
relevant phase space. These are hard to evaluate explicitly.} entering
in Eqs.~(\ref{general}) now take 
a modified form~\cite{rs2}\ $(Q^{\prime\,2}=-t)$, 
\begin{eqnarray}
\frac{d\hat{\sigma}_{L,\,\Sigma}^{\gamma^\ast g}}{d\,Q^{\prime \, 2}}&=&
\frac{8\pi^2\alpha_{\rm em}}{Q^2}\,\sum_q\,e_q^2
 \,\int\frac{d\xpr}{\xpr}\,
\frac{x}{\xpr}\,P_{L,\,\Sigma}\left(x,\xpr,\frac{Q^{\prime 2}}{Q^2}\right)\,
\sigma^{(I)}_{q^\ast g}(\xpr,Q^{\prime \, 2}), \label{sigma} \\[1ex]
\frac{d\sigma_{T}^{\gamma^\ast g}}{d\,Q^{\prime \, 2}}&=&
\frac{1}{2}\,\left(\frac{d\sigma_{\Sigma}^{\gamma^\ast g}}{d\,Q^{\prime \, 2}}
+\frac{d\sigma_{L}^{\gamma^\ast g}}{d\,Q^{\prime \, 2}}\right).
\label{planar}
\end{eqnarray}
The $\gamma^\ast\Rightarrow \overline{q}\,q$ ``flux'' factors~\cite{rs2,mrs-unpub},
\begin{eqnarray}
 P_{{\rm L} \atop{\rm \Sigma}}\left(x,\xpr,\frac{Q^{\prime 2}}{Q^2}\right)=\frac{3}{16\pi^3}\frac{x}{\xpr}\left\{\begin{array}{l}
2\, \frac{x}{x^{\prime}}\,
\frac{\Qpr}{Q^2}\,\left(1-\frac{\Qpr}{Q^2}\frac{x}{\xpr}\right)\,\frac{1-\xpr}{\xpr}
\\[1.5ex]
\left(1+\frac{1}{x}-\frac{1}{\xpr}-\frac{\Qpr}{Q^2}\right)
\end{array}\right., 
\end{eqnarray}
turn out to be directly related to the square of the pQCD photon wave function,
$\mid \Psi^{\rm pQCD}_{L,\,T}\mid^2$ as we shall see
explicitly below.
Corresponding to the more complex final state, Eqs.~(\ref{sigma}), (\ref{planar}) now
involve an additional integration over the Bjorken-$x^\prime$
variable $1\ge x^\prime\equiv Q^{\,\prime\,2}/(2p\cdot q^\prime)\ge x\ge 0$
of the $I$-induced subprocess, 
\begin{equation}
\left\{\overline{q}^{\,\ast} \mbox{\rm\ or\ }q^{\,\ast}\right\} \ (q^\prime) + g\ (p) \stackrel{(I)}{\Rightarrow} X, 
\label{subprocess}
\end{equation}
with total cross section $\sigma^{(I)}_{q^\ast
g}(\xpr,Q^{\prime\,2})$ that includes the main instanton dynamics
(see below).  

By means of a change of variables like in
Eqs.~(\ref{vartrans}), except for the replacement $x\Rightarrow
x/x^\prime$ due to $x^\prime\neq 1$, one now finds approximately
(assuming $z\gg (1-z)$  throughout without restriction),
\begin{eqnarray} 
\sigma_{L,\,T}(\xbj,Q^2)&\approx&\int dz \int
\frac{d^2\mathbf{l}}{(2\pi)^2} \mid  \tilde{\Psi}^{\rm
pQCD}_{L,\,T}(z,l)\mid^2\, \tilde{\sigma}^{(I)}_{\rm
dipole}(l,\xbj,\ldots);\hspace{3ex} \mbox{\rm with} \\[2ex]
\mid \tilde{\Psi}^{\rm pQCD}_{L\atop T}(z,l)\mid^2&=&\sum_q
e_q^2\frac{6\alpha_{\rm em}\,\hat{Q}^2}{(\hat{Q}^2+l^2)^2}\left\{\begin{array}{l}4\,z\,(1-z)\\
l^2/\hat{Q}^2\,(z^2+(1-z)^2)\end{array}\right.\\[1ex]
\tilde{\sigma}^{(I)}_{\rm dipole}(l,\xbj,\ldots) &\approx&
\xbj\,G(\xbj,\mu^2)\,\int_0^{\sqrt{s^{\,\prime}_{\rm max}}}
d\,E \left[\frac{\left((p+q^\prime)^2\right)^{3/2}}{4\,(p\cdot
q^\prime)\,Q^{\,\prime 2}}\, 
\sigma^{(I)}_{q^\ast \,g}\left(E,\frac{\hat{Q}^2+l^2}{z}\right)\right].
\label{sigdipglue}
\end{eqnarray} 
Since the total c.m. energy $\sqrt{s^{\,\prime}}$ of the
$q^\ast\,g\Rightarrow X$ subprocess (\ref{subprocess}) is given by
$\sqrt{s^{\,\prime}}=Q^\prime\,\sqrt{1/x^\prime -1}$, the $x^\prime$
integration above is equivalent to an integration over
$E\equiv\sqrt{s^{\,\prime}}$.  
The function $\tilde{\Psi}^{\rm pQCD}_{L,\,T}(z,l)$ is just
the $2d$-Fourier transform (cf. Eq.~(\ref{2dFT1})) of  $\Psi^{\rm 
pQCD}_{L,\,T}(z,r)$ in Eq.~(\ref{wavefu}). By inserting the known results for
$\sigma^{(I)}_{q^\ast g}$ from
Ref.~\cite{rs2} into  Eq.~(\ref{sigdipglue}), one
finds the following structure for $\tilde{\sigma}^{(I)}_{\rm
dipole}(l,\xbj,\ldots)$,
\begin{eqnarray}
\frac{d\tilde{\sigma}^{(I)}_{\rm
dipole}}{dE}
&\approx& \frac{1}{\alpha_{\rm s}}\,\xbj\,G(\xbj,\mu^2)\,\frac{\pi^5}{6}\,
\int_0^\infty d\,\rho\,D(\rho)\, \rho^5\,(\rho\,Q^\prime){\rm K_1}(\rho\,Q^\prime)
\int_0^\infty d\,\bar{\rho}\,D(\bar{\rho})\,\bar{\rho}^5\,(\bar{\rho}\,Q^\prime){\rm
K_1}(\bar{\rho}\,Q^\prime)\,\nonumber\\
&&\times\int \frac{d^4 R}{(\rho\bar{\rho})^{3/2}}\,
e^{i\,(p+q^\prime)\cdot R}\,\int dU
\,e^{-\frac{4\pi}{\alpha_{\rm s}}\,\Omega_{\rm
valley}\left(\xi\left(\frac{R^2}{\rho\bar{\rho}},\frac{\rho}{\bar{\rho}}\right),U\right)}\left\{\ldots
\right\};\hspace{3ex}  \sqrt{(p+q^\prime)^2}=E. 
\label{sigdip}
\end{eqnarray}
For reasons of space, we have skipped in $\{\ldots\}$ some (flavour
dependent) prefactors of secondary importance. 
The second line in Eq.~(\ref{sigdip}) is largely associated with the
final-state gluons. Let us briefly recall some of the essential features.  

While in case of the simplest $I$-induced process (\ref{simplest})
above, the contribution to the total cross section was obtained by
explicitly squaring the scattering amplitude 
and integrating over the final-state phase space, the derivation of
the DIS results~\cite{rs2} for the inclusive process (\ref{dis}) was
based on the optical theorem combined with the $I\bar{I}$-valley
method~\cite{yung}. In this 
approach~\cite{optvalley,valley-most-attr-orient,rs2}, one most
efficiently evaluates the total cross section from 
the imaginary part of the forward elastic amplitude
induced by the $I\bar{I}$-valley background $A_\mu^{(I\bar{I})}$.
This method elegantly accounts  for a resummation
and exponentiation of the final-state gluons, whose effects are
encoded in the explicitly known $I\bar{I}$-valley interaction~\cite{valley-most-attr-orient,valley-gen-orient}, 
\begin{equation}
\Omega_{\rm valley}(\xi,U)=S^{(I\bar{I})}_{\rm valley}(\xi,U)-1 =
\frac{\alpha_{\rm s}}{4\pi}\,S[A_\mu^{(I\bar{I})}] -1,
\end{equation}
appearing in Eq.~(\ref{sigdip}).
Apart from its dependence on the relative $I\bar{I}$-orientation $U$
in colour space, the valley action is restricted by conformal invariance to
depend only on the dimensionless, ``conformal separation''
\begin{equation}
\xi\equiv\frac{-R^2+i\epsilon R_0}{\rho\bar{\rho}}
+\frac{\rho}{\bar{\rho}} +\frac{\bar{\rho}}{\rho},
\label{xi}
\end{equation}
where in Euclidean space, the collective coordinate $R^{(E)}_{\mu}$
denotes the $I\bar{I}$-distance 4-vector, with $-R^2 \Rightarrow
R^{2\,(E)}\ge 0$ such that $\xi^{(E)}\ge 2$.  

In principle, the next step is to transform
Eq.~(\ref{sigdip}) further into the $(\mathbf{r},z)$ colour-dipole
representation, in generalization of Eq.~(\ref{final}). To this end, 
however, we first have to locate any possible, additional $l^2=l^2(Q^{\,\prime\,2},x/x^\prime,Q^2)$ dependences that might arise from the
final-state gluons etc., i.e. from the second line in
Eq.~(\ref{sigdip}). Let us begin by exhibiting a number of important
features of $d\tilde{\sigma}^{(I)}_{\rm dipole}/dE$ in
Eq.~(\ref{sigdip}) that emerge in the  softer $Q^{\,\prime}$ regime in combination with lattice results.   

\begin{figure} 
\begin{center}
\parbox{5.75cm}{\includegraphics*[width=5.75cm]{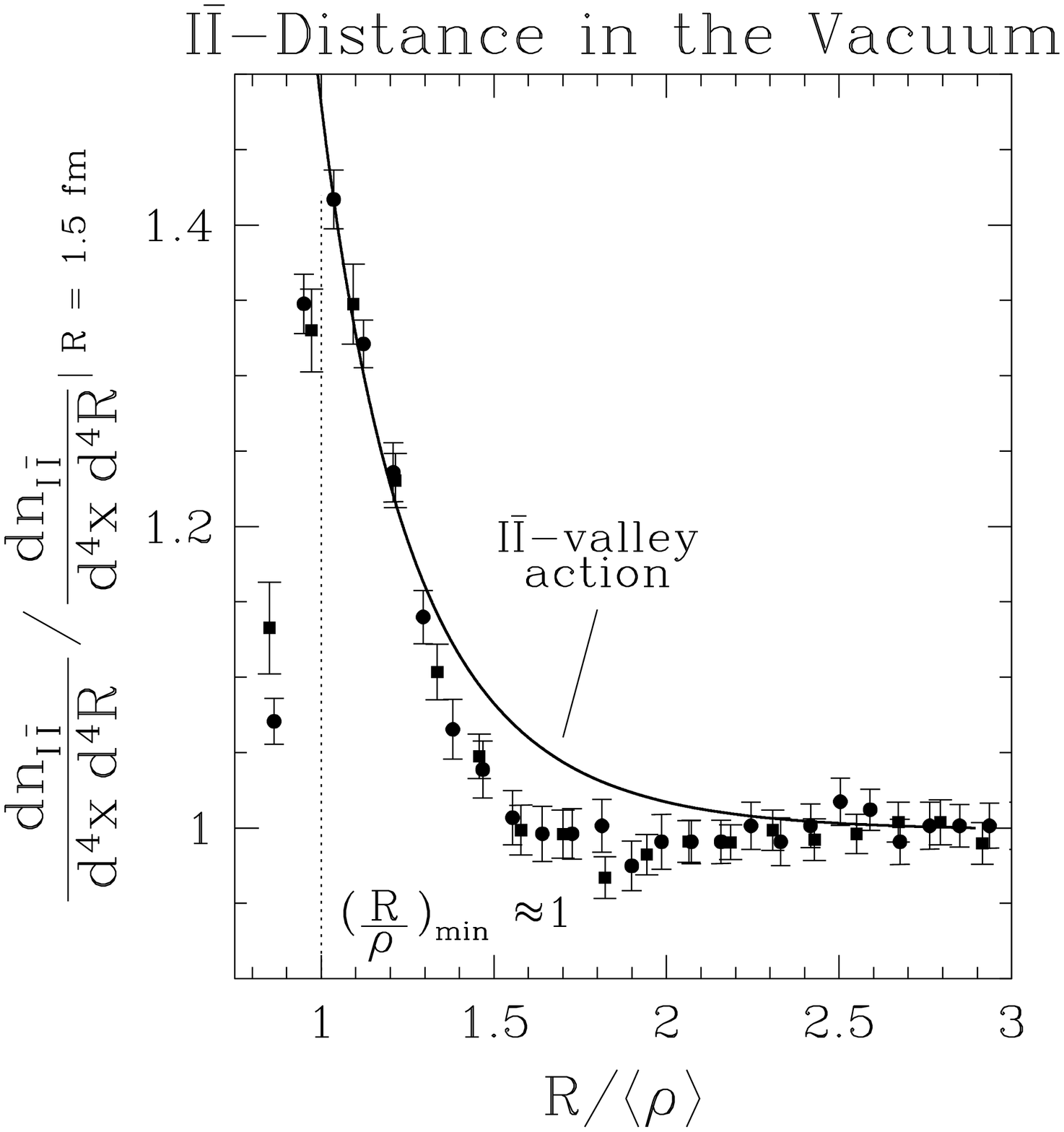}}\hfill
\parbox{5.49cm}{\vspace{-0.15cm}\includegraphics*[width=5.49cm]{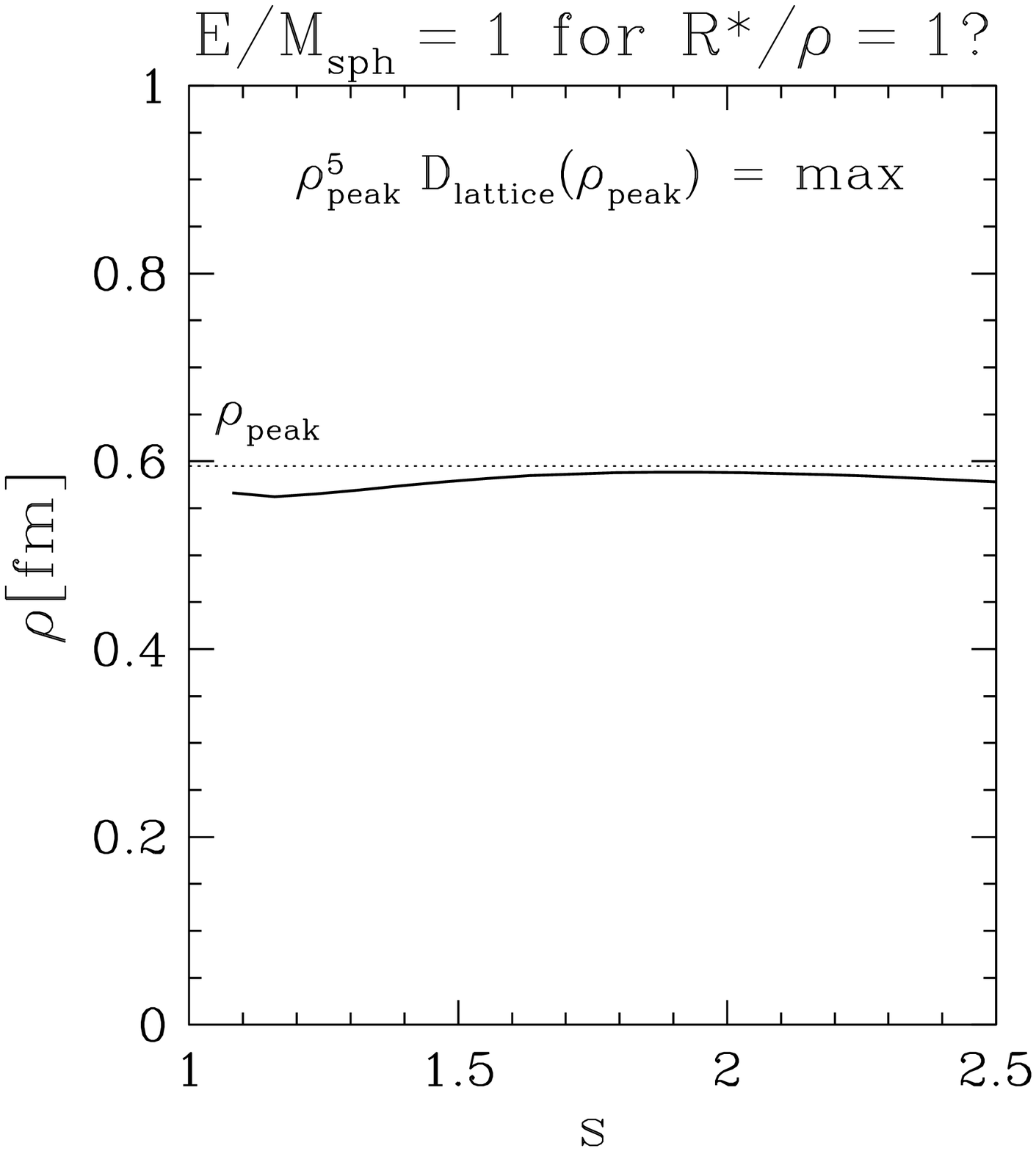}}\hfill
\parbox{5.75cm}{\includegraphics*[width=5.75cm]{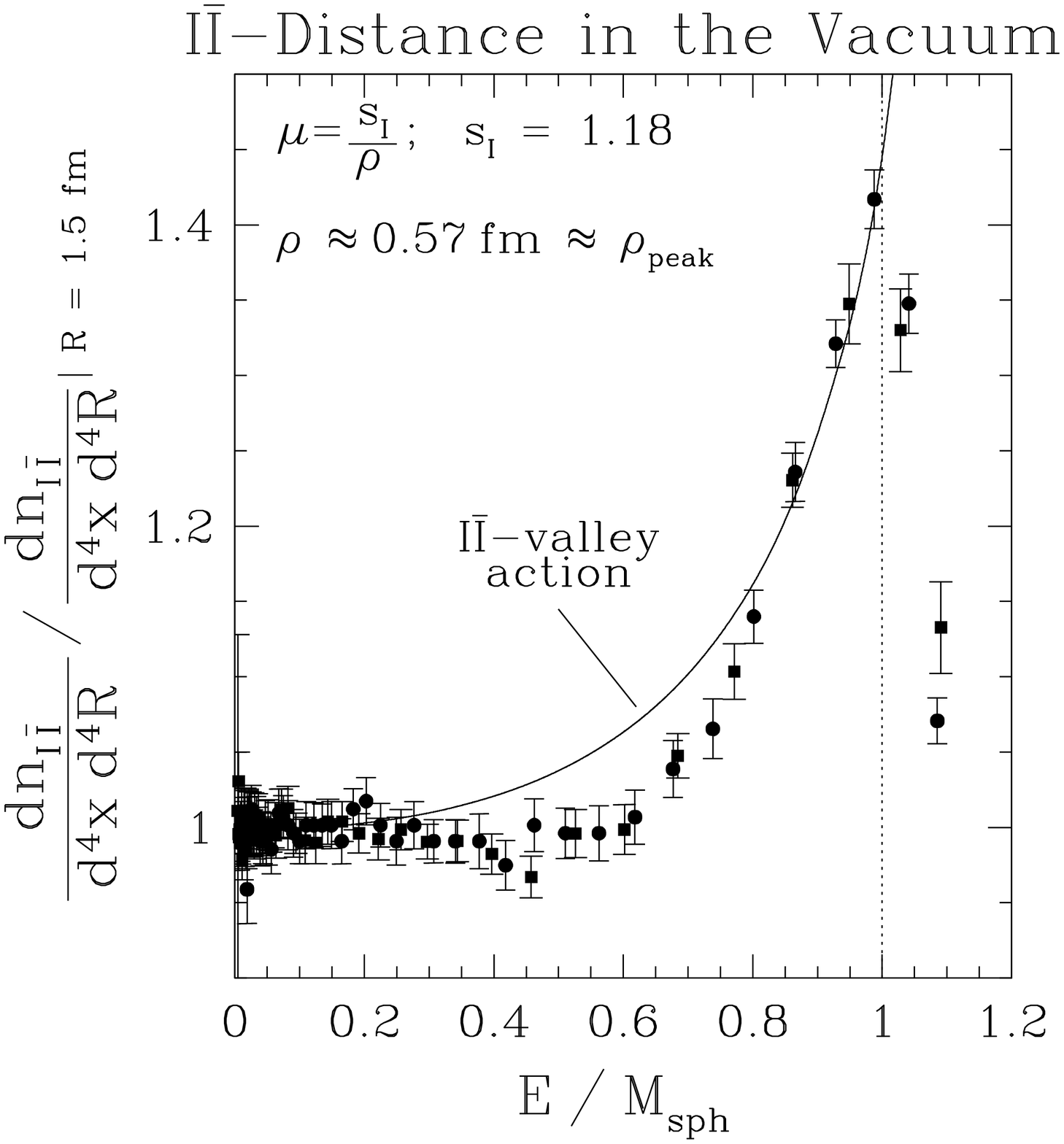}}
 \caption[dum]{\small\label{pic2} 
(Left) UKQCD lattice
 data~\cite{ukqcd,rs-lat,rs3}  of the (normalized) $I\bar{I}$-distance
 distribution versus the $I\bar{I}$-distance $R$ in units of
 $\langle\rho\rangle$ for quenched QCD ($n_f = 0$). The
 $I\bar{I}$-valley approximation appears to be 
reliable down to $R/\langle\rho\rangle\approx 1$, where it breaks down
abruptly. 
(Middle) 
If $\rho\approx \rho_{\rm peak}$, as is the case in
Eq.~(\ref{sigdip}) towards soft $Q^\prime$, the saddle-point relation
(\ref{xi-saddle}) associates $E/M_{\rm sph}=1$ with $R^\ast/\rho=1$.
The weak s-dependence
signals approximate renormalization group invariance ($\mu_r=s/\rho$ ).
(Right) The \mbox{$I\bar{I}$-distance}
 distribution, being largely a measure
of $\left\langle \exp{\left[-\frac{4\pi}{\alpha_{\rm s}(s/\rho_{\rm
peak})}\, \Omega_{\rm valley}(\xi,U)\right]}\right\rangle_{U}$ in
Eq.~(\ref{sigdip}), displayed 
versus energy in units of the QCD sphaleron mass $M_{\rm
 sph}$. While the valley prediction continues to rise for
 $E/M_{\rm sph}>1$, the lattice data provide the first direct
 evidence that the $I\bar{I}$-valley approach is adequate
 right up to $E\approx M_{\rm sph}$, where the dominant contribution to the
 scattering process arises.}
\end{center}
\end{figure}

Besides the $I$-size distribution $D(\rho)$,  the
$I\bar{I}$-interaction $\Omega$ in Eq.~(\ref{sigdip}) represents a
second crucial quantity of the $I$-calculus, for which we shall exploit
independent lattice information that will be instrumental for a
transition towards softer $Q^\prime$. Fig.~\ref{pic2} (left) displays
(normalized) UKQCD lattice 
data~\cite{ukqcd,rs-lat,rs3} of the $I\bar{I}$-distance distribution
versus the (Euclidean) $I\bar{I}$-distance 
$R \equiv \sqrt{R^{\,2\,(E)}}$ in units of 
 $\langle\rho\rangle$ for quenched QCD ($n_f = 0$), along with the
 prediction of the $I\bar{I}$-valley approach~\cite{rs-lat}, 
\begin{equation}
\frac{dn^{\rm valley}_{I\bar{I}}}{d^4\,x\,d^4\,R}=\int_0^\infty d\rho\,
D_{\rm lattice}(\rho)\,\int_0^\infty d\bar{\rho}\, D_{\rm
lattice}(\bar{\rho})\,\int d\,U \,
e^{-\frac{4\pi}{\alpha_{\rm s}(s/\sqrt{\rho\bar{\rho}})}\,\Omega_{\rm valley}\left(\xi\left(\frac{R^2}{\rho\bar{\rho}},\frac{\rho}{\bar{\rho}}\right),U\right)}.
\label{lattice}
\end{equation}  
Note the remarkable similarity in structure of this lattice
``observable'' and $d\tilde{\sigma}^{(I)}_{\rm dipole}/dE$ in 
Eq.~(\ref{sigdip}). This holds  notably in the soft $Q^\prime$ regime where the
exponential suppression of larger size instantons via the ${\rm K}_1$
Bessel functions in Eq.~(\ref{sigdip}) tends to vanish, i.\,e. $(\rho
Q^\prime)\,{\rm K}_1(\rho Q^\prime)\sim 1$, and instead $\rho \approx
\bar{\rho}\approx \rho_{\rm 
peak}\approx \langle\rho\rangle$, with $\rho_{\rm peak}$ and
$\langle\rho\rangle$ being the (close-by) positions of the {\em sharp}
peaks of $\rho^5\,D_{\rm lattice}(\rho)$ and $D_{\rm lattice}(\rho)$,
respectively (cf. Fig.~\ref{pic} (left)).

Indeed, Fig.~\ref{pic2} (left) reveals crucial information
concerning the range of validity of the $I\bar{I}$-valley interaction
$\Omega_{\rm valley}$. The $I\bar{I}$-valley
approximation appears to be quite reliable down to
\mbox{$\left(\frac{R}{\langle\rho\rangle}\right)_{\rm min}\approx 1$}, where
the $I\bar{I}$-distribution shows a sharp peak, while the valley
prediction continues to rise indefinitely.
According to Eq.~(\ref{xi}), with
$\rho\approx\bar{\rho}$, this peak of the lattice data 
corresponds to $\xi_{\rm peak} \approx 3$ and hence to  
$S^{(I\bar{I})}_{\rm valley}(\xi_{\rm
peak}=3,U^\ast)\approx\frac{1}{2}$, for the 
most attractive $I\bar{I}$ colour orientation $U=U^\ast$ that is known
to dominate the U-integral in Eqs.~(\ref{sigdip}) at least for sufficiently 
 large values of $4\pi/\alpha_{\rm s}$ in form of a saddle
 point. This important result perfectly matches with
 previous theoretical claims~\cite{uni,diak-petrov}, according to
 which the maximal $I$-induced (QCD or EW) cross section shows a
 ``square-root'' enhancement compared to the pure tunneling behaviour
 at $E=0$ \mbox{($S^{(I\bar{I})}_{\rm valley}(\xi=\infty,U^\ast)=1$)}. 
\eject
Let us demonstrate next that this marked peak of the
lattice $I\bar{I}$-distance distribution in Fig.~\ref{pic2} (left) 
in fact corresponds to the top of the potential barrier, i.\,e. to the
sphaleron mass, \mbox{$E\approx M_{\rm sph}$}, which may be
estimated~\cite{diak-petrov} as the potential energy of the instanton
field exactly in the middle of the transition when the instanton
passes the  $N_{\rm Chern-Simons}=1/2$ point,      
\begin{equation}
M_{\rm sph}(\rho)=\frac{1}{g_{\rm s}^2}\,\frac{1}{4}\,4\pi\,\int
dr\,r^2\,\frac{96\rho^4}{(t^2+r^2+\rho^2)^4}_{\mid
t=0}=\frac{3\pi}{4}\,\frac{1}{\alpha_{\rm s}\,\rho}.
\label{sphaleron}
\end{equation}
This result for $M_{\rm sph}$ matches with the  estimate  $M_{\rm
sph}\sim Q^{\,\prime}$  from Ref.~\cite{rs1} at large $Q^{\,\prime}$,
where the integrals in Eq.~(\ref{sigdip}) are known to be dominated by
a unique saddle-point in all integration variables, notably including 
$\rho=\bar{\rho}\approx\rho^\ast(Q^{\,\prime})\sim 1/(\alpha_{\rm
s}\,Q^{\,\prime})$. 

For {\em large} $Q^\prime$, the familiar saddle point~\cite{rs2}, 
\begin{equation}
\rho^\ast
=\bar{\rho}^{\,\ast}\sim \frac{1}{\alpha_{\rm s}\,Q^{\,\prime}};\hspace{2ex} 
R^{\,\ast}_\mu=(-i\rho^\ast\sqrt{\xi^\ast(x^\prime)-2},\
\mathbf{0}), 
\end{equation}
of the 
effective exponent $\Gamma$ in Eq.~(\ref{sigdip}) is determined by
requiring $\Gamma$ to be stationary with respect to {\em all} integration
variables. In particular, the combination of  
$\partial\Gamma/\partial\xi=0$ and $\partial\Gamma/\partial\rho=0$ 
leads to a unique solution\footnote{Taking for
simplicity the additional saddle-point relations
$\rho^\ast=\bar{\rho}^{\,\ast},\ \mathbf{R^\ast}=0,\ U=U^\ast$ for
granted already.}
$\xi^\ast=\xi^\ast(x^\prime,\ldots)$ for {\em all} physical values of
$x^\prime$, $x\le x^\prime\le 1$. 

However, the situation changes
drastically, in the softer $Q^\prime$-regime, where $\rho\approx
\rho_{\rm peak}\approx 0.59$ fm  with 
$\rho_{\rm peak}$ corresponding to the {\em sharp} peak position of
$\rho^5\,D_{\rm lattice}(\rho)$ in
Eq.~(\ref{sigdip}). Here, effectively only
$\partial\Gamma/\partial\xi=0$ remains 
and provides together with Eq.~(\ref{sphaleron}), a correlation of
$E/M_{\rm sph}(\rho)$ and $\xi^\ast=2+R^{\,2\,(E)\,\ast}/\rho^{\,2}$ for
$\rho\Rightarrow \rho_{\rm peak}$. At 2-loop renormalization group
accuracy, we obtain from Eq.~(17) of Ref.~\cite{rs2}, with
renormalization scale $\mu=s/\rho$ and $s=\mathcal{O}(1)$
(e.\,g. $s_I\approx 1.18$, cf. Ref.~\cite{rs-lat}),    
\begin{equation}
\frac{E}{M_{\rm sph}}={\frac {32}{3}}\, \frac{d\,\Omega_{\rm valley}(\xi^\ast)}{d\xi} \,
\sqrt {{\xi^\ast}-2}\, \left( 1-\frac{1}{2\,\pi}\,\ln (s)\,\alpha_{\rm s}\left( {\frac {s}{{\rho}}} \right) {\beta_0}-\frac{1}{
8\,\pi^2}\,\ln (s)\, \alpha_{\rm s} \left( {\frac {s}{
{\rho}}} \right)^2{\beta_1}\right). 
\label{xi-saddle}
\end{equation}
First of all, we notice from Eq.~(\ref{xi-saddle}) that for soft
$Q^\prime$, a saddle-point solution 
for $\xi^\ast$ only exists if $E/M_{\rm sph}$ is not too large. The
reason is that $\frac {32}{3}\, \frac{d\,\Omega_{\rm valley}(\xi^\ast)}{d\xi} 
\sqrt {{\xi^\ast}-2} \,\lwig\, 3.5$, with the maximum attained around
$\xi^\ast\approx 2.4$, i.\,e. quite near to the striking peak position
$\xi_{\rm peak}\approx 3$ of the lattice data for the $I\bar{I}$-distance
distribution above. Hence,  it is 
tempting to ask, for which values of $\rho$ and the scheme parameter
$s$ the peak value $\xi_{\rm peak}=3$  would exactly correspond to $E=M_{\rm
sph}$. The solution from Eq.~(\ref{xi-saddle}) with a 3-loop expression for
$\alpha_{\rm s}$ and $\Lambda^{(\rm nf=0)}_{\rm \overline{MS}}=238$ MeV
from the lattice~\cite{alpha} is displayed in
Figs.~\ref{pic2} (middle), (right) and nicely confirms our intuitive
expectations.  

{\em In summary:} For soft $Q^{\,\prime}$, i.\,e. $\rho\approx\rho_{\rm
peak}$ and {\em increasing} total energy E of the $I$-subprocess
(\ref{subprocess}), $0 \lwig E <M_{\rm sph}(\rho_{\rm peak})$, the (Euclidean)
saddle-point solution $\xi^\ast$ of Eq.~(\ref{xi-saddle}) {\em
decreases} such that      
$d\tilde{\sigma}^{(I)}_{\rm dipole}/dE$ in 
Eq.~(\ref{sigdip}) steeply {\em increases} until a sharp maximum is reached. 
Fig.~\ref{pic2} (right) illustrates this behaviour by displaying
instead the $I\bar{I}$-distance distribution that is largely a measure
of $\left\langle \exp{\left[-\frac{4\pi}{\alpha_{\rm s}(s/\rho_{\rm
peak})}\, \Omega_{\rm valley}(\xi,U)\right]}\right\rangle_{U}$,
versus $E/M_{\rm sph}$ from 
Eq.~(\ref{xi-saddle}). Fig.~\ref{pic2} (middle) shows that the maximum
position $R\approx \rho_{\rm peak}$, as inferred 
from lattice data,  indeed corresponds to the top of the potential
barrier, i.e. to $E\approx M_{\rm sph}$, provided $Q^{\,\prime}$ approaches the
soft regime and thus stirs $\rho,\bar{\rho}$ towards $\rho_{\rm peak}$
in Eq.~(\ref{sigdip}). For $E>M_{\rm sph}$ the Euclidean saddle point
\mbox{$R_0=-i\rho^\ast\sqrt{\xi(E/M_{\rm sph}) -2}$}, described by
Eq.~(\ref{xi-saddle}), ceases to exist and $d\tilde{\sigma}^{(I)}_{\rm
dipole}/dE$ may be estimated from the peaking of $\left\langle \exp{\left[-\frac{4\pi}{\alpha_{\rm s}(s/\rho_{\rm peak})}\, \Omega_{\rm
valley}(\xi,U)\right]}\right\rangle_{U}$ (lattice)  
to decrease again in this regime. Finally, from the lattice data, the
underlying $I\bar{I}$-valley approximation has been found to
interpolate reliably between the pure 
tunneling regime ($E=0$) and the sphaleron at the top of the
potential barrier ($E=M_{\rm sph}$). Altogether, the resulting picture
is in qualitative agreement with the findings of Refs.~\cite{uni,shuryak11}. 

In view of the above analysis, the integration over the total
$I$-subprocess energy $E$ in 
Eqs.~(\ref{sigdipglue}), (\ref{sigdip}) up to $E=\sqrt{s^{\,\prime}_{\rm
max}}$, may evidently be extended to 
$E\Rightarrow\infty$ due to the strong peaking of  $d\tilde{\sigma}^{(I)}_{\rm 
dipole}/dE$ around $E\approx M_{\rm sph}(\rho_{\rm peak}) <
\sqrt{s^{\,\prime}_{\rm max}}$, 
\begin{eqnarray}
\lefteqn{\tilde{\sigma}^{(I)}_{\rm dipole}(l,\ldots) \approx\int_0^\infty
dE\;\frac{d\tilde{\sigma}^{(I)}_{\rm dipole}}{dE}
\approx \frac{1}{\alpha_{\rm
s}}\,\xbj\,G(\xbj,\mu^2)\, \frac{\pi^5}{6}}
\nonumber\\[1ex] &&\times  \label{gluedip}
\int_0^\infty d\rho\,D(\rho)\,\rho^5\,(\rho\,Q^\prime){\rm K_1}(\rho\,Q^\prime)
\int_0^\infty d\bar{\rho}\,D(\bar{\rho})\,\bar{\rho}^5\,(\bar{\rho}\,Q^\prime){\rm
K_1}(\bar{\rho}\,Q^\prime)\,H_{\rm sph}(\rho,\bar{\rho}),
\end{eqnarray} 
with the dimensionless function $H_{\rm sph}(\rho,\bar{\rho})$, being
largely associated with the final-state gluons\footnote{In
Eq.~(\ref{gluons}), the notation $\Omega_{\rm
lattice}\left(\xi\left(\frac{R^2}{\rho\bar{\rho}},\frac{\rho}{\bar{\rho}}\right),U\right)$
is meant to denote the $I\bar{I}$-valley interaction for $E\lwig M_{\rm
sph}$, supplemented by the additional constraints for $E>M_{\rm
sph}$ from the lattice data, as discussed above (cf. Fig.~\ref{pic2}).
},  
\begin{equation}
H_{\rm sph}(\rho,\bar{\rho})\approx\int_0^\infty dE\,
\int \frac{d^4 R}{(\rho\bar{\rho})^{3/2}}\,
e^{i\,(p+q^\prime)\cdot R}\int d\,U
\;e^{-\frac{4\pi}{\alpha_{\rm s}(s/\sqrt{\rho\bar{\rho}})}\,\Omega_{\rm
lattice}\left(\xi\left(\frac{R^2}{\rho\bar{\rho}},\frac{\rho}{\bar{\rho}}\right),U\right)}\left\{\ldots
\right\}.
\label{gluons}
\end{equation}
In the soft $Q^\prime$ regime, $H_{\rm sph}(\rho,\bar{\rho})$ does not
introduce any additional $l$-dependences beyond those coming 
from the ``master integrals'' $\mathcal{R}\left((\hat{Q}^2+l^2)/z\right)$ in
Eq.~(\ref{gluedip}) in analogy to Eq.~(\ref{masterI}) in case of the
simplest $I$-induced process. Hence we may perform the $2d$-Fourier
transformation $(\mathbf{l}\Rightarrow \mathbf{r})$ and finally obtain
(for $z \gg (1-z)$ without restriction) e.\,g., 
 \begin{eqnarray}
\lefteqn{\left(\left|\Psi_L\right|^2\sigma_{\rm dipole}\right)^{(I)}
 \approx\, \mid\Psi_L^{\rm pQCD}(z,r)\mid^{\,2}\,
\frac{1}{\alpha_{\rm s}}\,\xbj\, G(\xbj,\mu^2)\,\frac{\pi^5}{6}}\nonumber\\[1ex]
&&\times\int_0^\infty\,d\rho\, D(\rho)\,\rho^5\,
\int_0^\infty\, d\bar{\rho}\, D(\bar{\rho})\,\bar{\rho}^{\,5}\,
H_{\rm sph}(\rho,\bar{\rho})\,
\frac{-\frac{d}{dr^2}\left(2 r^2
\frac{\mbox{K}_1(\hat{Q}\sqrt{r^2+\rho^2/z})}{\hat{Q}\sqrt{r^2+\rho^2/z}
}\right)\times (\rho\leftrightarrow\bar{\rho})}{{\rm K}_0(\hat{Q}r)^2}. 
\end{eqnarray}
For $r\gwig\rav$,  
  \begin{equation}
 \left(\left|\Psi_L\right|^2\sigma_{\rm dipole}\right)^{(I)}
\approx\,
\mid\Psi_L^{\rm pQCD}(z,r)\mid^{\,2}\,\sigma_{\rm dipole}^{(I)\,{\rm gluons}},
\end{equation}
with
\begin{equation}
 \sigma_{\rm dipole}^{(I)\,{\rm gluons}}=
\frac{1}{\alpha_{\rm s}}\,\xbj\, G(\xbj,\mu^2)\,\frac{\pi^5}{6}
\left(\int_0^\infty\, d\rho\,D_{\rm lattice}(\rho)\,\rho^5 \right)^2
H_{\rm sph}(\rav,\rav).
\end{equation}
Similar to the simplest $I$-induced process (\ref{final}), the result
exhibits a saturating, geometrical limit, proportional to the area
$\pi\,\mathcal{R}(0)^2=\pi\left(\int_0^\infty\,d\rho\,D_{\rm
lattice}(\rho)\,\rho^5\right)^2$, subtended by the instanton.

{\em Outlook:}
An investigation of the phenomenology associated
with the emerging picture of soft high-energy processes induced by
instantons is challenging and in progress~\cite{su3}. Before more
quantitative predictions can be made, a careful study of inherent
uncertainties are necessary. Let us merely state at this point that
the instanton-induced contributions indeed appear significant towards
the soft regime. Like in case of
the extensively studied DIS processes (HERA) induced by {\em small}
instantons (cf. e.\,g. Refs.~\cite{fs,h1-final}), one expects characteristic
final-state signatures. Given the importance of
lattice data for the conclusions reached in this paper, further
improved lattice results in this direction would be most desirable.     
While the main intention of this paper was to associate the
origin of the conspicuous geometrical scale in diffractive scattering
with the average instanton size, clearly, a number of important aspects remain
to be investigated. For instance, an understanding of the
mechanism that causes the cross section to increase with energy in
an instanton framework is of importance.   

\vspace{2ex}
\noindent
{\em Acknowledgements:} 
We are grateful to Leonid Frankfurt and Mark Strikman for valuable
discussions  and thank Andreas Ringwald for a careful
reading of the manuscript.

\end{document}